\documentclass{article}
\usepackage{spconf,amsmath,graphicx}

\usepackage[table]{xcolor}
\usepackage{amssymb}
\usepackage{multirow}

\title{Speaker conditioning of acoustic models using affine transformation for multi-speaker speech recognition}
\name{Midia Yousefi, John H.L. Hansen}
\address{  Center for Robust Speech Systems (CRSS), Erik Jonsson School of Engineering,\\
University of Texas at Dallas, Richardson, Texas, USA}
\begin{document}
%
\maketitle
\begin{abstract}

This study addresses the problem of single-channel Automatic Speech Recognition of a target speaker within an overlap speech scenario. In the proposed method, the hidden representations in the acoustic model are modulated by speaker auxiliary  information to recognize only the desired speaker. Affine transformation layers are inserted into the acoustic model network to integrate speaker information with the acoustic features. The speaker conditioning process allows the acoustic model to perform computation in the context of target-speaker auxiliary information. The proposed speaker conditioning method is a general approach and can be applied to any acoustic model architecture. Here, we employ speaker conditioning on a ResNet acoustic model. Experiments on the WSJ corpus show that the proposed speaker conditioning method is an effective solution to fuse speaker auxiliary information with acoustic features for multi-speaker speech recognition, achieving +9\% and +20\% relative WER reduction for clean and overlap speech scenarios, respectively, compared to the original ResNet acoustic model baseline. 

\end{abstract}
\section{Introduction}
\label{sec:intro}


Multi-talker speech recognition is focused on recognizing individual speech sources from overlap speech, and is one main challenge for current ASR systems \cite{chang2020end,watanabe2020chime,yousefi21_interspeech, yousefi2018assessing,barker2018fifth,qian2018past,mirsamadi2019multi,yousefi2020block}. 
Current solutions for multi-speaker speech recognition  can be categorized into two main approaches: \emph{(i)} performing front-end speech processing based on separation on the overlap speech, then applying ASR to the separated speech signals \cite{boeddeker2018front,yousefi2016supervised,yousefi2020frame,deng2011front, narayanan2014investigation,mirsamadi2014multichannel, yousefi2019probabilistic}; or \emph{(ii)} skipping the explicit separation step and developing a multi-speaker speech recognition system directly using either hybrid \cite{kanda2019acoustic,weng2015deep,kanda2019simultaneous} or end-to-end \cite{seki2018purely,lu2021streaming} ASR frameworks. Recently, an end-to-end multi-speaker speech recognition system was proposed based on Transformers \cite{chang2020end}. This approach achieved considerable improvement at the expense of more computational cost for a reasonable temporal resolution. In another study \cite{chen2017progressive}, overlap speech was considered as a mismatch condition of the clean speech recognition scenario, and teacher-student training was employed for transfer learning from clean to overlap speech. The main drawback of this approach is requiring training sets with parallel clean and overlapped speech, which is difficult to collect in real-world applications \cite{denisov2019end}.
Recently, several studies \cite{kanda2019acoustic,wang2019end,subramanian2020far} have used speaker-specific embeddings to learn a frame-level mask for the target speaker which suppresses interfering speech. Although these approaches use the additional speaker-specific information to guide the ASR system, their main limitation is that they assume only one speaker is active in each Time-Frequency bin.

To address the challenges of single-channel multi-speaker speech recognition, in this study, we focus on speaker conditioning of the Acoustic Model (AM) by performing an affine transformation.
In contrast to former approaches which employ speaker embedding to estimate speaker-specific masks, we propose to use speaker embedding to compute parameters of the affine transformation, allowing the acoustic model to conduct its computation in the context of the desired speaker auxiliary information. The proposed speaker conditioning method is a general approach and can be applied to any AM architecture. In this paper, we employ  speaker conditioning on a ResNet acoustic model in hybrid DNN-HMM setup. Experiments are performed on WSJ corpus, achieving +9\% and +20\% relative WER reduction for clean and overlap speech scenarios, compared to the original ResNet acoustic model. The contributions of this paper are threefold:
\begin{itemize}
    \item Proposing speaker conditioning of the ResNet acoustic model using an affine transformation.
    \item Comparing the proposed method with alternate feature-wise  acoustic model transformations such as conditional biasing and middle feature-map fusion.
    \item Evaluating the performance of the proposed speaker conditioned ASR trained on an alternate input feature called Wav2Vec representation. 
\end{itemize}

The remainder of this paper is organized as follows. In Sec.\ref{sec:sys}, the problem is outlined and proposed method described. Sec.\ref{sec:exp}, presents experiments and results. Finally the conclusions are discussed in Sec.\ref{sec:con}. 

\begin{figure*}[t]
\centering
\vspace{-0.9cm}
\includegraphics[width=\linewidth]{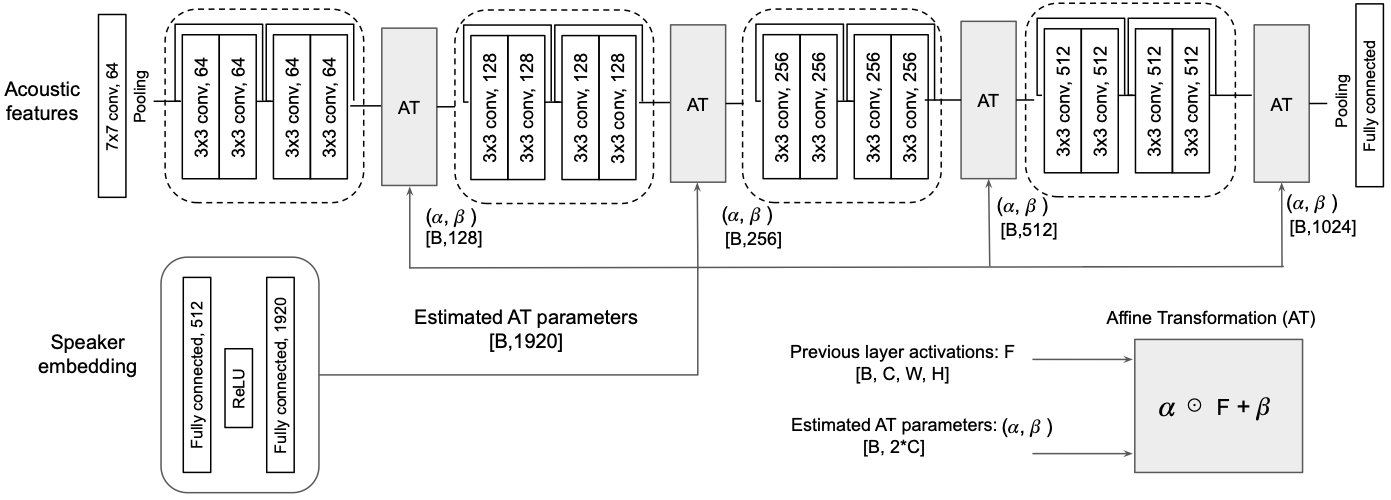}
\vspace{-0.7cm}
\caption{The proposed speaker conditioned ResNet18 acoustic model using Affine Transformation (AT) blocks.}
\label{fig:model}
\vspace{-0.2cm}
\end{figure*}

\section{Single-channel multi-speaker ASR}
\label{sec:sys}

Multi-speaker speech recognition can gain substantial improvement by deploying other sources of information such as speaker identity in addition to acoustic features \cite{denisov2019end, wang2019end}. However, designing an efficient method to fuse a combination of multiple sources (i.e., acoustic features and speaker embeddings) to obtain higher quality and improved information is still a challenging task. Additionally, capturing complex interactions between multiple sources should maintain a balanced compromise between model/network computational cost and performance. A popular method to address this problem is to use a feature-wise transformation \cite{perez2018film} which can model the complex relation between speaker-specific characteristics and acoustic features in a multi-speaker speech scenario to identify and recognize the desired speaker in the mixed speech recording. This transformation can be performed in several manners such as \emph{conditional biasing}, \emph{conditional scaling}, and \emph{conditional affine transformation}. In this section, we focus on a conditional affine transformation  which is a more general approach. The aforementioned conditional biasing and scaling are two specific examples of conditional affine transformation. 

\subsection{Conditional affine transformation} Affine Transformation (AT) influences the output of the acoustic model network by applying a linear modulation to the network's intermediate features. This modulation is parameterized by shifts and coefficients obtained based on speaker-specific embedding. Let $x$ be a context-expanded window of acoustic features for overlap speech, and $y_s$ be a phoneme label or a senone alignment (i.e., from GMM-HMM) for the target-speaker speech signal. DNN acoustic models are used estimate the posterior probability as:
\begin{equation}
   p(y_s|x, s) = DNN(x, z_s),
\end{equation}
where DNN is typically trained to maximize the log probability of the phoneme alignment or minimize the cross-entropy error, and $s$ is the target speaker with an x-vector \cite{snyder2018x} embedding $z_s$. In this study, the original ResNet18 model is considered as our baseline. Next, affine transformation layers are inserted into ResNet18 network to build the speaker conditioned acoustic model. The scale and bias factors of AT are  estimated by a two-layer fully connected network $h$ based on x-vector $z_s$ as:
\begin{equation}
    (\alpha_{i,c},\; \beta_{i,c}) = h(z_s)
\end{equation}
where  $i$ and $c$ refer to the $i$-th data sample in the minibatch, and the $c$-th channel feature map. Once $\alpha_{i,c}$ and $\beta_{i,c}$ are estimated, they are used to modulate the ResNet's intermediate activations $F_{i,c}$ as:
\begin{equation}
      AT(F_{i,c}^{l}| \alpha_{i,c}, \beta_{i,c}, F_{i,c}^{l-1}) = \alpha_{i,c} \odot  F_{i,c}^{l-1} + \beta_{i,c}
\end{equation}
where $AT$ and $l$ represent the Affine Transformation, and network's layer. The proposed speaker conditioned ResNet18 is shown in Fig.\ref{fig:model}.  The speaker embedding x-vector is submitted to the network $h$ to estimate a $[B, 1920]$ matrix which is $(\alpha_{i,c}$, $\beta_{i,c})$ pairs of AT layers. Each AT layer receives two inputs:  the previous layer output, and the $(\alpha_{i,c}$, $\beta_{i,c})$ pair. The dimension of $\alpha_{i,c}$ and $\beta_{i,c}$ is $[B, C]$ each. In the AT layer, each channel of the extracted feature map is scaled by $\alpha_{i,c}$ and shifted by $\beta_{i,c}$ to modulate the feature-map distribution of activations based on the target-speaker embedding.

\begin{table}[h]
\vspace{-0.3cm}
\caption{Comparing our ResNet18 acoustic model baseline with other approaches on WSJ (WER in \%).}
\vspace{-0.3cm}
\begin{minipage}{0.45\textwidth}
\centering
\begin{tabular}{@{\extracolsep{0.1pt}}lc c c} 
\\\hline 
\hline \\
\textit{System}  & \multicolumn{1}{c}{Dev-93} & \multicolumn{1}{c}{Eval-92} \\ 
\hline 
Lee et al. 2021 \cite{lee2021intermediate} &  12.0 & 9.9\\
Higuchi et al. 2020 \cite{higuchi2020mask} &15.4 &	12.1\\
Chi et al. 2020 \cite{chi2020align} & 13.7&	11.4\\
Rouhe et al. 2020 \cite{rouhe2020speaker} & 13.2 &	9.3\\
Sabour et al. 2018 \cite{sabour2018optimal} &- &	9.3\\
Borgholt et al. 2020 \cite{borgholt2020end} &- &	9.3\\
Park et al. 2019 \cite{lee2019simple} & - &	7.8\\
\hline
Our baseline & 12.1	& 7.9

\end{tabular} 
\end{minipage}
\vspace{-0.5cm}
\label{tab:base}
\end{table}

\begin{table*}[t]
\setlength{\extrarowheight}{4pt}
\begin{center}
\vspace{-0.4cm}
\caption{ WER of the proposed speaker conditioned ResNet18 acoustic model with Affine Transformation (AT) in different settings. Each experiment is repeated three times and the average WER is reported.}

\begin{tabular}{ |c|c|c|c|c|c|c|c|c|c| }
\cline{2-10}

\multicolumn{1}{c|}{} & \multicolumn{6}{c|}{Simulated overlap speech based on Dev-93} &\multicolumn{3}{c|}{Clean speech} \\ 
\cline{3-6} \cline{7-9} 
\hline
Acoustic model & 0dB & 5dB & 10dB & 15dB & 20dB & 25dB & Dev-93 & Eval-92 & Eval-93 \\
\hline
ResNet18 (baseline) & 65.06&58.29&47.03&34.72&24.36&17.62&12.14&7.92&10.81 \\
ResNet18 + AT (proposed) &63.83&	55.83&	43.30&	29.90&	20.34&	15.15&	11.50&	7.66&	9.64\\
\hline
ResNet18 + AT (bias=0) &63.69&	55.23&	43.57&	30.15&	19.95&	15.19&	11.75&	7.66&	9.57\\
ResNet18 + AT (scale=1)& 65.10& 57.66&	46.39&	32.79&	22.29&	16.65&	12.25&	7.91& 10.57\\
ResNet18 + AT (sigmoid(scale))& 64.63&	56.66&	45.33&	31.78&	21.680&	16.250&	11.993&	7.803&	10.263\\
ResNet18 + AT (tanh(scale))& 64.53&	56.82&	45.27&	31.64&	21.19&	16.02&	11.99&	7.72&	9.97\\
\hline
ResNet18 + AT (Block1)& 63.68&	55.33&	\textbf{43.31}&29.44&	\textbf{19.51}&\textbf{14.67}&11.56&\textbf{7.49}&	\textbf{9.85}\\
ResNet18 + AT (Block 1-2)&\textbf{63.50}&	\textbf{55.19}&	43.60&	\textbf{29.33}&	19.79&	14.85&	\textbf{11.33}&	7.50&	9.88\\
ResNet18 + AT (Block 1-3)&63.51&	55.29&	43.35&	29.65&	19.87&	14.88&	11.49&	7.55&	9.93\\
ResNet18 + AT (Block 4)& 64.66&	57.66&	47.28&	33.86&	23.27&	16.54&	11.64&	8.19&	10.47\\
\hline
\end{tabular}
\vspace{-0.4cm}
\label{tab:at}
\end{center}
\end{table*}


\section{Experiments and results}
\label{sec:exp}
In this section, we investigate the performance of the proposed speaker conditioning method presented in Fig.\ref{fig:model} on WSJ corpus. 
In order to conduct the experiments, clean \textit{tr-si284} is used in the training phase for all acoustic models. We partitioned \textit{tr-si284} into a training set $(90\%)$ and a held-out cross-validation set $(10\%)$. ASR performance for different acoustic models are reported in terms of Word-Error-Rate (WER) on clean \textit{dev-93}, \textit{eval-93}, and \textit{eval-92}. Additionally, overlap speech is generated based on \textit{dev-93} by selecting random utterances from random speakers and adding them with Signal-to-Interference Ratio (SIR) ranging from $0$ to $25$dB with increments of $5$dB. The baseline acoustic model is ResNet18 with $3400$ output senones.  The  network parameters  are  updated  by  the  gradients  of the cross entropy loss using Stochastic Gradient Descent (SGD) optimizer with a momentum of $0.9$ and initial learning rate $0.01$. The training process is completed by performing early stopping \cite{zhang2016understanding}. The maximum number of epochs is set to $100$, batch size  $1024$ context-expanded frames; learning rate is decreased by $50\%$ if the cv loss improvement is less than $0.01$ for $3$ successive epochs. The early stopping is performed if no improvement is observed on the cv loss once the learning rate has decayed $6$ times. 13-dim MFCC computed over a $25ms$ window with $10ms$ shift with a $20$ frame context ($10$ frames on each side) is used for training the acoustic model. Consistent with the standard Kaldi recipe for WSJ, we use the trigram language models provided by LDC for WSJ data. In order to minimize the effect of parameter initialization on the acoustic model and final WER, we repeat each experiment three times with different initial parameters.

\begin{table*}[t]
\setlength{\extrarowheight}{4pt}
\begin{center}
\vspace{-0.6cm}
\caption{ WER of the proposed speaker conditioned ResNet18 based on Affine Transformation (AT) compared to other fusion techniques. Each experiment is repeated three times and the average WER is reported.}
\begin{tabular}{ |c|c|c|c|c|c|c|c|c|c| }

\cline{2-10}

\multicolumn{1}{c|}{} & \multicolumn{6}{c|}{Simulated overlap speech based on Dev-93} &\multicolumn{3}{c|}{Clean speech} \\ 
\cline{3-6} \cline{7-9} 
\hline
Signal-to-Interference Ratio & 0dB & 5dB & 10dB & 15dB & 20dB & 25dB & Dev-93 & Eval-92 & Eval-93 \\
\hline
ResNet18 + Conditional biasing  & 64.66&	57.20&	45.74&	32.75&	23.54&	17.87&	13.10&	8.44&	11.22\\
ResNet18 + Middle fusion & 63.94&	57.51&	47.81&	34.02&	23.19&	16.68&	11.81&	8.24&	10.64 \\
\textbf{ResNet18 + AT (proposed)}& \textbf{63.68}&	\textbf{55.33}&	\textbf{43.31}&\textbf{29.44}&	\textbf{19.51}&\textbf{14.67}&\textbf{11.56}&\textbf{7.49}&	\textbf{9.85}\\
\hline
\end{tabular}
\label{tab:com}
\end{center}
\end{table*}

\begin{figure*}
\centering
\begin{tabular}{cc}
\includegraphics[height=3.8cm,width=7cm]{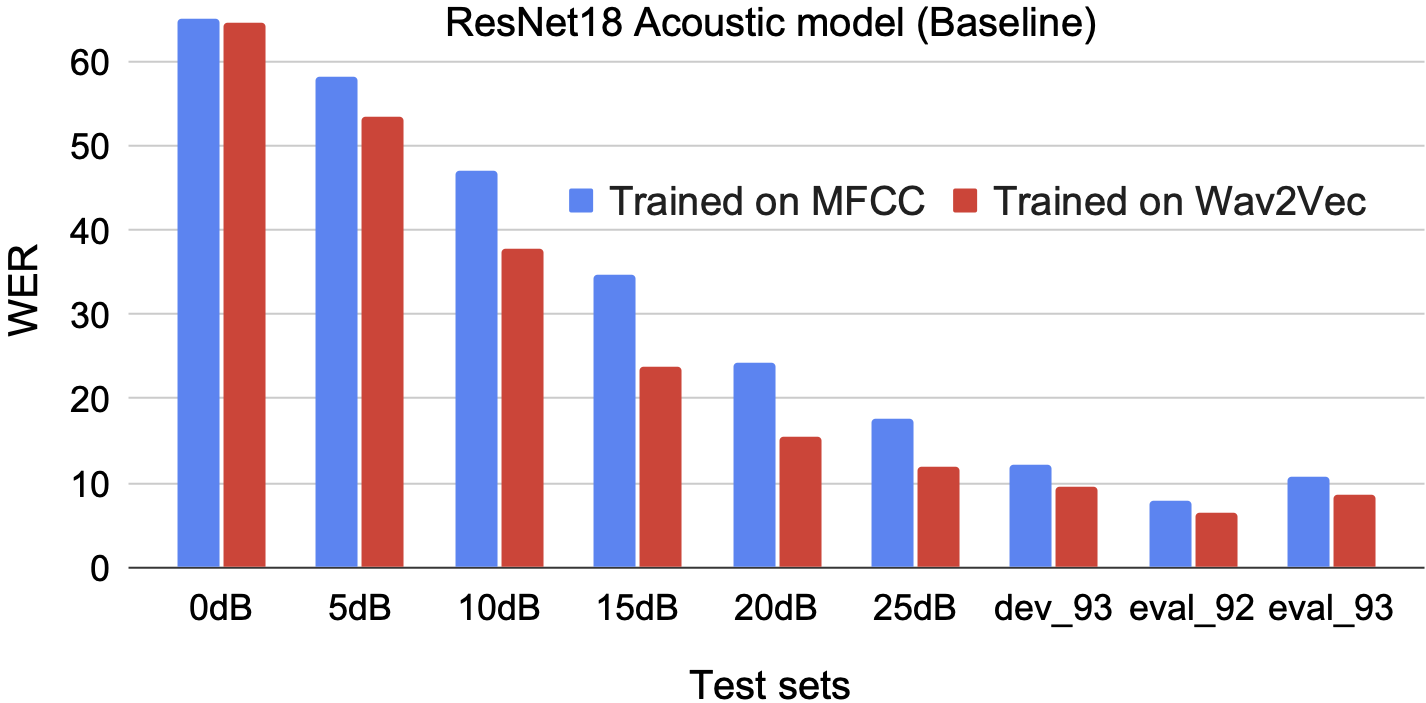}&
\hspace{1.5cm}
\includegraphics[height=3.8cm,width=7cm]{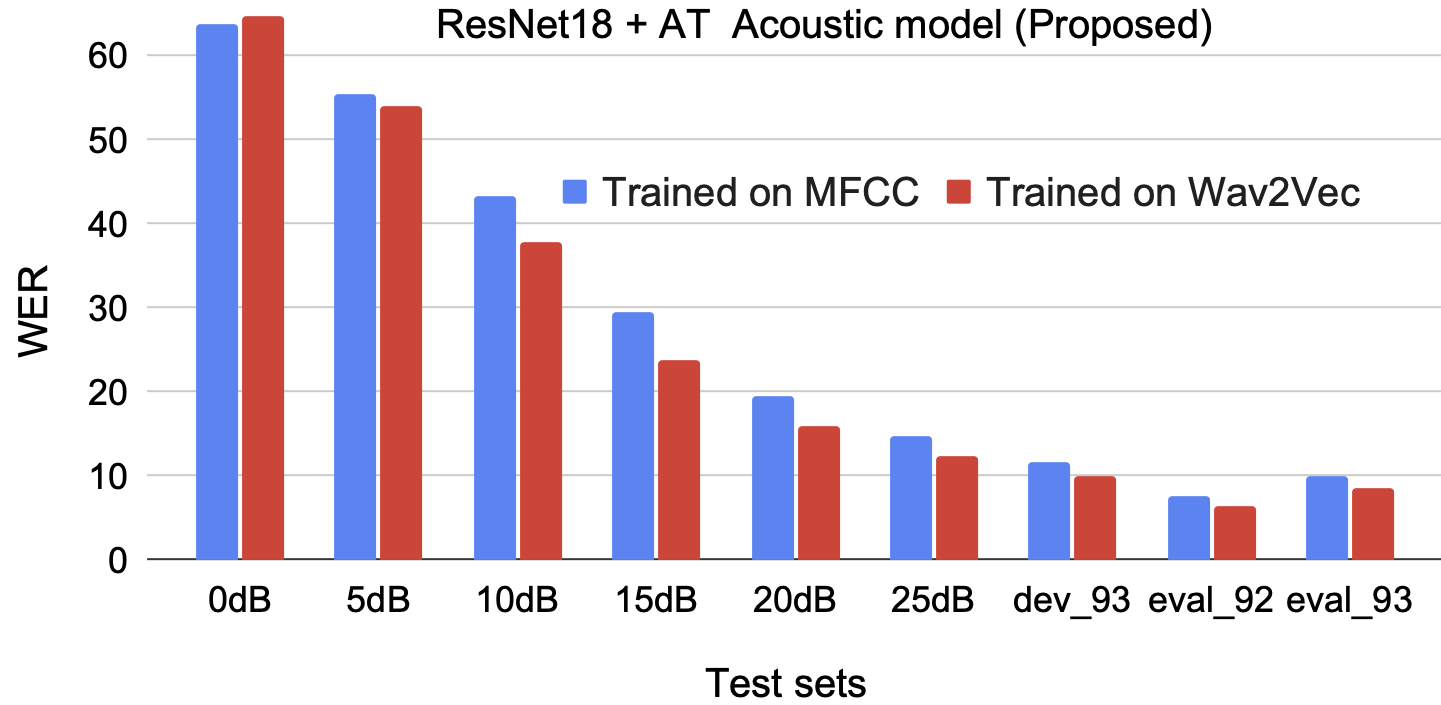} 
\end{tabular}
\vspace{-0.2cm}
\label{figur}\caption{ WER of ResNet18 baseline and proposed ResNet18 + AT trained on MFCC and Wav2Vec input features.}
\vspace{-0.2cm}
\label{fig:wav2vec}
\end{figure*}

Performance of the ResNet18 baseline is compared with recent studies in Table \ref{tab:base}. The main purpose of this comparison is to ensure that our baseline achieves a competitive performance compared to recent studies, and it is seen we have a strong starting point for further developing our proposed speaker conditioning technique.  There are other approaches that leverage transfer learning, semi supervised learning, or more advanced language models to achieve further improvement. However, since we focus on speaker conditioning of acoustic model, we train our ResNet18 acoustic model only on \textit{tr-si284}, and use Kaldi for training the language model. The first row of Table \ref{tab:at} presents performance of the baseline on overlap speech, which is severely degraded. Therefore, we build on the ResNet18 acoustic model baseline and apply our Affine Transformation (AT) layers as depicted in Fig.\ref{fig:model}. The results for speaker conditioned ResNet using AT are presented in the second row of Table \ref{tab:at} which shows $+2\%$ relative improvement for severe overlap speech recordings (i.e., $0$dB) and an average of $+5\%$ relative improvement on clean test sets. Since AT effectively performs speaker-adaptation, the trained acoustic model is tuned to the target speaker, therefore, it achieves better performance even on the clean test sets. The maximum relative improvement is achieved for input SIR $20\%$ in which the level of overlap speech is neither too severe nor too easy for the acoustic model; therefore, the  target-speaker auxiliary information can be very helpful in improving performance.  

Moreover, the effect of $\alpha$ and $\beta$ is studied separately by setting $\alpha=1$ and $\beta=0$. The result in Table \ref{tab:at} manifest that the effectiveness of the conditional Affine Transformation can be mainly attributed to the scale coefficient rather than the shift parameter. Therefore, we further investigate the effect of  $\alpha$ by restricting its value to $(0,1)$ using Sigmoid, and $(-1,1)$ using $\tanh$ function. Nevertheless, the results reported in Table \ref{tab:at} reveal that unrestricted $\alpha$ achieves better performance which may be due to the flexibility it provides for the network to learn the range that best suits the data. So far, the AT layers have been applied to all ResNet18 blocks (each dashed rectangular in Fig. \ref{fig:model} is considered as a block). To find the best network depth in which AT layers are most effective, several experiments are conducted with AT only applied to specific individual blocks. Based on these experiments, the AT layers are most effective when applied only to the first block (block1), and least effective when only applied to the last block (block4). However, applying  AT layers to the first two blocks (block1-2) and the first three blocks (block 1-3) did not improve ASR performance, while it differently adds computational cost. To summarize our findings based on the experiments, unrestricted-scale Affine Transformation applied to the initial blocks of the ResNet18 acoustic model achieves the best overall results while simultaneously maintaining the lowest computational cost. 

Next, the proposed method is compared with other speaker conditioning techniques in Table \ref{tab:com}. Conditional biasing refers to adding speaker information (x-vector) as a bias to the acoustic features in the first hidden layer. Middle fusion refers to adding the x-vector to the intermediate extracted feature map after the second block. Therefore, the intermediate feature map is conditioned before entering block 3 for extracting further higher-level features adapted to the target speaker. As shown in Table \ref{tab:com}, the proposed speaker conditioning based on Affine Transformation outperforms all other approaches in both clean and overlap speech scenarios. 

So far, the focus of this study has been on designing the acoustic model. However,  performance of the acoustic model can further improve by deploying more robust input features other than MFCC. In the final section, we evaluate the proposed method trained on noise-invariant Wav2Vec features \cite{schneider2019wav2vec}. Wav2Vec representation has been trained on large amounts of unlabeled audio data in an unsupervised manner. Fig. \ref{fig:wav2vec} (left) shows the WER of the baseline ResNet18 trained on MFCC and Wav2Vec features, which manifests the effectiveness of Wav2Vec in reducing the WER across all test sets in the absence of speaker auxiliary information. The highest improvement is achieved for overlap speech with SIR $15$dB, which is $+11\%$ absolute improvement in WER. Fig.\ref{fig:wav2vec} (right) depicts the WER of the proposed speaker conditioned ResNet18 trained on MFCC and Wav2Vec. Similar to the baseline, the speaker conditioned acoustic model benefits from the Wav2Vec features by achieving $+6\%$ absolute improvement in WER for SIR $15$dB. However, due to the availability of speaker information, the acoustic model is less sensitive to the robustness of the input acoustic features, and thus, the amount of improvement from Wav2Vec is less in the proposed speaker conditioned ResNet18 compared to the baseline. In conclusion, the WER across all test sets is improved by using the proposed speaker conditioned acoustic model trained on wav2Vec. For example, on the overlap speech test set with SIR $15$dB, the proposed ResNet18 with Affine Transformation trained on Wav2Vec gains +33\% relative (+11\% absolute) improvement in WER compared to the original ResNet trained on MFCC.

\section{Conclusion}
\label{sec:con}

In this study, we proposed a speaker conditioning method for acoustic modeling in multi-speaker speech recognition. In the proposed method, Affine Transformation layers are inserted into the acoustic model architecture to fuse speaker-specific information with the acoustic model. The proposed speaker conditioned acoustic model  was compared with  other fusion techniques such as early fusion of speaker embedding and middle feature-map fusion. Additionally, the performance of the proposed method was evaluated on alternate input features called Wav2Vec. The results on WSJ corpus clearly demonstrate that the proposed speaker conditioned acoustic model based on affine transformation achieves consistent WER improvement for clean and overlap speech scenarios.




%

\bibliographystyle{IEEEbib}
\bibliography{refs}
\end{document}